\documentclass[aps,prl,twocolumn,groupedaddress]{revtex4}

\usepackage{graphicx}
\usepackage{dcolumn}
\usepackage{bm}

\begin{document}

\preprint{APS/123-QED}

\title{Ultrasound Attenuation of Superfluid $^3$He in Aerogel}

\author{H.C. Choi}
\author{N. Masuhara}
\author{B.H. Moon}
\author{P. Bhupathi}
\author{M.W. Meisel}
\author{Y. Lee}
\email[]{yoonslee@phys.ufl.edu}
\affiliation{Microkelvin
Laboratory, Department of Physics, University of Florida,
Gainesville, FL 32611-8440, USA}
\author{N. Mulders}
\affiliation{Department of Physics and Astronomy, University of
Delaware, Newark, DE 19716, USA}
\author{S. Higashitani}
\author{M. Miura}
\author{K. Nagai}
\affiliation{Faculty of IAS, Hiroshima University, Kagamiyama
1-7-1, Higashi-Hiroshima 739-8521, Japan}
\date{\today}

\begin{abstract}
We have performed longitudinal ultrasound (9.5 MHz) attenuation
measurements in the B-phase of superfluid $^3$He in 98\% porosity
aerogel down to the zero temperature limit for a wide range of
pressures at zero magnetic field. The absolute attenuation was
determined by direct transmission of sound pulses. Compared to the
bulk fluid, our results revealed a drastically different behavior
in attenuation, which is consistent with theoretical accounts with
gapless excitations and a collision drag effect.
\end{abstract}

\maketitle

Liquid $^{3}$He has attracted intense interest for many decades in
the field of low temperature physics\,\cite{vol}.  In its normal
state, liquid $^{3}$He has served as a paradigm for a Fermi liquid
whose nature transcends $^3$He physics. The superfluid phases of
$^{3}$He exhibit exotic and intriguing features associated with
the broken symmetries in the condensate, having an unconventional
structure of the order parameter with spin triplet {\it p-wave}
pairing. Liquid $^{3}$He is arguably the most well-understood
system mainly because of its extreme intrinsic pureness at low
temperatures. Therefore, it has provided important insights in
understanding other unconventional superconductors such as the
high temperature superconductors, the heavy fermion
superconductors, and in particular the more recently discovered
Sr$_{2}$RuO$_{4}$, which is also thought to have the {\it p-wave}
symmetry\,\cite{APmac}. However, the same virtue has hampered the
effort in pursuing answers to an important overarching question:
how does the nature of a quantum condensate (spin triplet {\it
p-wave} superfluid in this case) respond to increasing impurity or
disorder?

Observation of superfluid transitions in liquid $^3$He impregnated
in high porosity aerogel in 1995\,\cite{por,spr95} opened a novel
path to introducing static disorder in liquid $^3$He. Aerogel
possesses a unique structure, whose topology is at the antipode of
widely studied porous media such as Vycor glass and metallic
sinters. Due to its open structure, there are no well-defined
pores in aerogel and consequently, the liquid is in the proximity
to the bulk. Ninety eight percent porosity aerogel, which has been
used in most of the studies including this work, offers a
correlated network of strand-like aggregates of SiO$_{2}$
molecules whose structure can be characterized by the geometrical
mean free path ($\ell \simeq$ 100 - 200 nm), the diameter of
strand ($r \approx$ 3 nm), and the average inter-strand distance
($d \simeq$ 25 - 40~nm).  The coherence length of pure superfluid
$^3$He, $\xi_0$, which varies from 20 nm (34 bar) to 80 nm (0
bar), is at least an order of magnitude larger than the strand
diameter but is comparable to $\ell$ and $d$. As a result, the
scattering off the aerogel strand would have a significant
influence on the superfluid. It is now well established that the
superfluid transition temperature is significantly depressed from
that of the bulk, and the effect of pair-breaking is progressively
magnified at lower pressures, leading to the possibility of a
quantum phase transition at $P_{c}\approx$~6 bars\,\cite{mat}. To
date, three distinct superfluid phases have been experimentally
identified, namely the A-like, B-like, and A$_{1}$-like
phases\,\cite{spr95,spr96,all,bark00,choi}. The B-like phase and
the A$_{1}$-like phase in aerogel show striking similarity to
their counterparts in the bulk superfluid\,\cite{dmi02,choi}.
Detailed NMR studies\,\cite{bark00,dmi02,all} suggest that the
aerogel B-phase has the same order parameter structure as the bulk
B-phase. The aerogel A$_{1}$-phase only appears in the presence of
magnetic field as is the case in the bulk \cite{choi}. However,
the aerogel A-phase exhibits quite a different behavior from the
bulk A-phase ({\it e.g.} in NMR frequency shift and superfluid
density), although the overwhelming experimental evidence suggests
that it is an equal spin pairing state. Various interpretations or
novel propositions on the possible order parameter structure have
been suggested for this phase\,\cite{volo,fom,vic}.

Nuclear magnetic resonance and ultrasound spectroscopy have been
used in concert to investigate the microscopic structure of the
superfluid phases\,\cite{vol,hal}. These two experimental methods
encompass complementary information on the orbital (ultrasound)
and spin (NMR) structure of the Cooper pairs. Rich spectra of
order parameter collective modes in bulk superfluids, which are
the fingerprints of specific broken symmetries in the system, have
been mapped by ultrasound spectroscopic techniques\,\cite{hal}. In
2000, Nomura {\it et al.}\,\cite{nom} performed ultrasound
attenuation measurements on 98\% aerogel using a 16.5~MHz cw
acoustic impedance technique. Their work was limited to a single
pressure at 16 bars and down to 0.6~mK. Although their technique
was not adequate in determining absolute attenuation, they managed
to extract the absolute sound attenuation after making auxiliary
assumptions.  A Bayreuth group\,\cite{hri} performed absolute
sound attenuation measurements in aerogel (97\% porosity) using a
direct sound transmission technique at 10~MHz.  They experienced
poor transducer response, and observed self-heating and no
depression in the aerogel superfluid transition. We conducted high
frequency sound transmission experiments in 98\% porosity aerogel,
covering the whole phase diagram of the superfluid phases in
aerogel, from 8 to 34~bars and from the transition temperatures to
as low as 200~$\mu$K.

In this experiment, two matched LiNbO$_{3}$ longitudinal sound
transducers with the fundamental resonance at 9.5~MHz were used as
a transmitter and a receiver. The 6.3~mm diameter transducers were
separated by a Macor spacer maintaining a 3.05~($\pm$~0.02)~mm
sound path between the transducers where the aerogel sample was
grown {\it in situ}. This scheme ensures the best contact between
the transducer surface and the aerogel, which is crucial for clean
sound transmission at the boundaries. A 1~$\mu$s pulse was
generated by the transmitter and detected by the receiver.
Temperature was determined by a melting pressure thermometer (MPT)
for $T\geq$~1~mK and a Pt NMR thermometer for $T\leq$~1~mK which
was calibrated against the MPT. No non-linear response or
self-heating was observed at the excitation level used in this
work.  All the data presented here, except for 8 bars, were taken
while warming with a typical warming rate of 3~$\mu$K/min. A
detailed description on the experimental cell and experimental
techniques can be found elsewhere\,\cite{choi2,choi3}.

\begin{figure}
\includegraphics[height=2.75in, angle=-90]{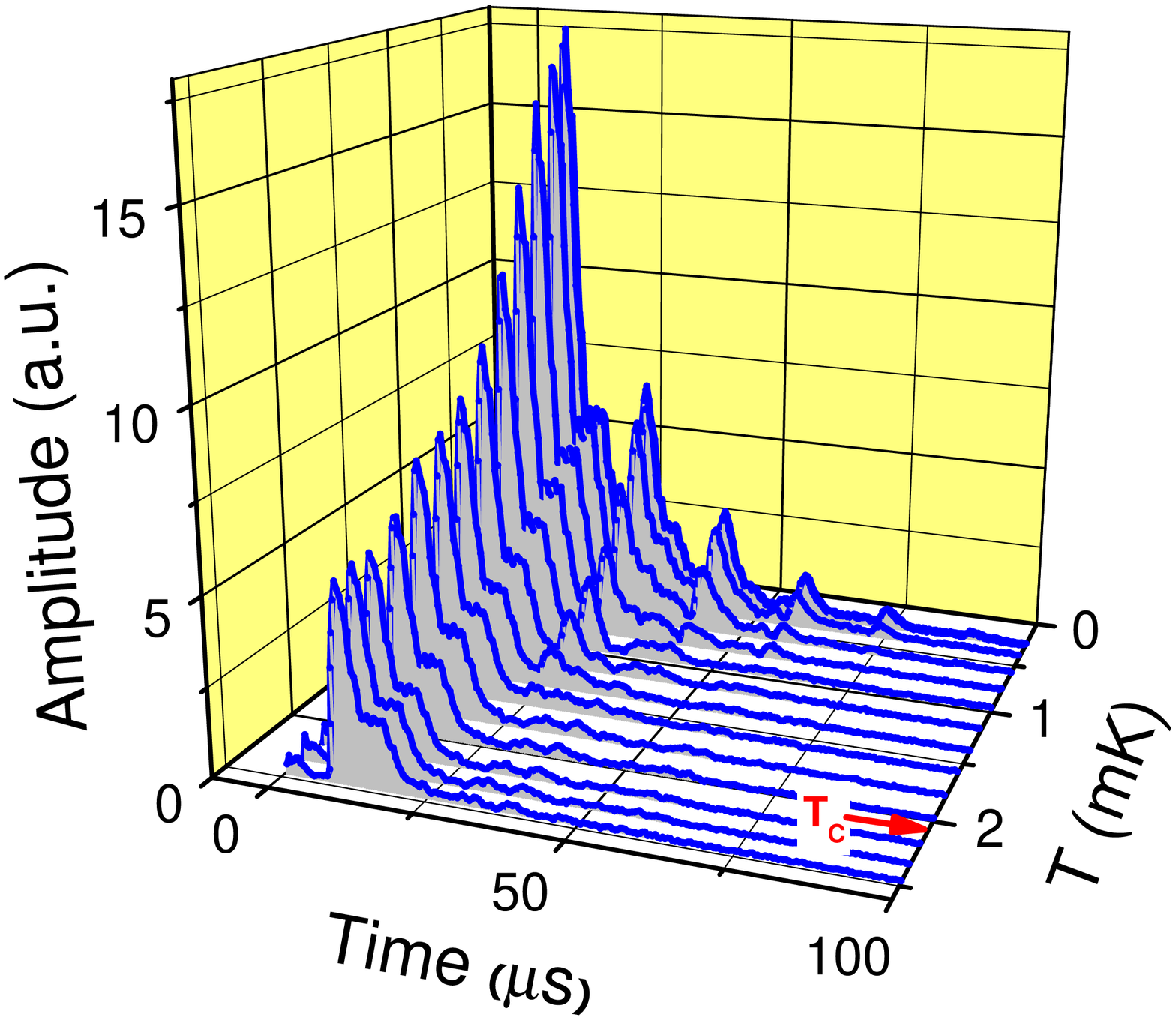}
\caption{\label{fig:epsart} Acoustic response from the receiver
vs. time at 34 bars for select temperatures ranging from 0.3~mK to
2.5~mK. The aerogel superfluid transition is marked by a small
arrow.}
\end{figure}

The temporal responses of the receiver taken at 34~bars are shown
in Fig.~1 for select temperatures ranging from 0.3 to 2.5~mK. The
primary response, which starts to rise around 8~$\mu$s, shows a
rather broad response due to ringing of the high Q transducer (Q
$\sim 10^3$).  The step-like structure of the receiver signal is
caused by the slight mismatch in the spectra of the
transducers\,\cite{choi3}. Below the aerogel superfluid transition
(marked around 2.1~mK by an arrow in Fig.~1) the primary response
starts to grow and the trailing echoes emerge from the background,
as the sound attenuation decreases in the superfluid. No change in
the receiver signal was observed at the bulk superfluid
transition. The multiple echoes follow a {\it bona fide}
exponential decay in time. Absolute sound attenuation was obtained
in the following manner\,\cite{lee07}. First, the relative
attenuation at each temperature was calculated using the area
under the primary response curve by integrating the signal from
the rising edge to a fixed point in time (23~$\mu$s point).  The
absolute attenuation at 0.4~mK and 29 bars, obtained using the
primary signal and the echoes, was used as a reference point in
converting the relative attenuation into the absolute attenuation.
Due to a drastic mismatch in the acoustic impedance at the the
transducer-aerogel/$^3$He boundary, the signal absorbtion at the
surface of transducers was ignored\,\cite{lee07}. The possible
background contributions to attenuation from the quasi-particle
scattering off the cavity wall\,\cite{esk} and the non-parallel
alignment of the two transducers are estimated to be negligible.

The absolute attenuations on warming for several pressures are
plotted as a function of temperature in Fig.~2(a). The superfluid
transition is marked by the smooth drop in attenuation. Our
aerogel superfluid transition temperatures are in excellent
agreement with the previously reported values for all
pressures\,\cite{mat,ger}. At 9.5 MHz in the {\it bulk} B-phase, a
strong attenuation peak appears right below the superfluid
transition. This peak is the result of the combined contributions
from pair-breaking and coupling to the order parameter collective
modes. Above the polycritical pressure, the B to A transition on
warming is registered as a sharp step in attenuation.  In {\it
aerogel}, none of these features exist.  However, we did observe a
sharp step in attenuation on cooling for $P >$~14 bars, which
implies the existence of the supercooled A-phase\,\cite{lee07}. We
were able to identify a rather smooth B to A transition on warming
for 29 and 34 bars within $\approx$ 150~$\mu$K below the
superfluid transition. This observation is consistent with the
previous results obtained using a transverse acoustic impedance
technique\,\cite{vic}. Therefore, most of the attenuation data
presented here are in the aerogel B-phase. In the bulk B-phase
with a clean gap, the attenuation follows $\alpha \propto
e^{-\Delta(T)/k_{B}T}$ below the attenuation peak, practically
reaching zero attenuation below $T/T_{c}\approx$~0.6, due to
thermally activated quasi-particles, where $\Delta(T)$ is the
temperature dependent gap and $k_{B}$ is the Boltzmann constant.
In contrast, the attenuation in aerogel decreases rather slowly
with temperature and remains high even at $T/T_{c} \approx$~0.2.
Furthermore, a peculiar shoulder feature appears at $T/T_{c}
\approx$~0.6 for higher pressures.  This feature weakens gradually
and eventually disappears at lower pressures, Fig.~2(a).

Sound propagation for higher harmonics up to 96~MHz was measured
for several temperatures and pressures, but no evidence of sound
propagation was found above 30~MHz even at 0.3~mK, where the
lowest attenuation is expected. Below about 10~mK, the scattering
process is dominated by the temperature independent impurity
scattering off the aerogel, and at 9.5~MHz, $\omega\tau_{i}
\sim$~0.1 for all pressures where $\tau_{i}=\ell/v_{f}$ (see below
for $\ell$). Therefore, the sound mode should remain in the
hydrodynamic limit. This claim is bolstered by the observation of
the strong frequency dependence in attenuation and the absence of
a temperature dependence in the normal fluid attenuation
\cite{nom}. The coupling between the normal component of the
superfluid $^3$He and the mass of the elastic aerogel modifies the
conventional two-fluid hydrodynamic equations\,\cite{mac,gol}.
This consideration leads to two (slow and fast) longitudinal sound
modes with different sound speeds,
$c_{s}=c_{a}\sqrt{\rho_{s}\rho_{a}}/\rho$, and
$c_{f}=c_{1}\sqrt{{1+\rho_{a}\rho_{s}/\rho_{n}\rho}\over{1+\rho_{a}/\rho_{n}}}$.
Here, $c_{f(s)}$ represents the speed of the fast (slow) mode,
$\rho_{n(s)}$ is the normal fluid (superfluid) density
($\rho=\rho_{n}+\rho_{s}$), $\rho_{a}$ is the aerogel density,
$c_{1}$ is the speed of hydrodynamic sound in $^3$He, and finally
$c_{a}$ is the sound speed of the bare aerogel. From the time of
flight measurements, we found the sound speed in aerogel
consistently lower (by $\approx$~20\%) than $c_{1}$ for all
pressures studied and in good agreement with the values obtained
using the expression above \cite{com1}. Detailed analysis of sound
velocity for various pressures will be presented in a separate
publication.

\begin{figure}
\includegraphics[height=2.75in, angle=-90]{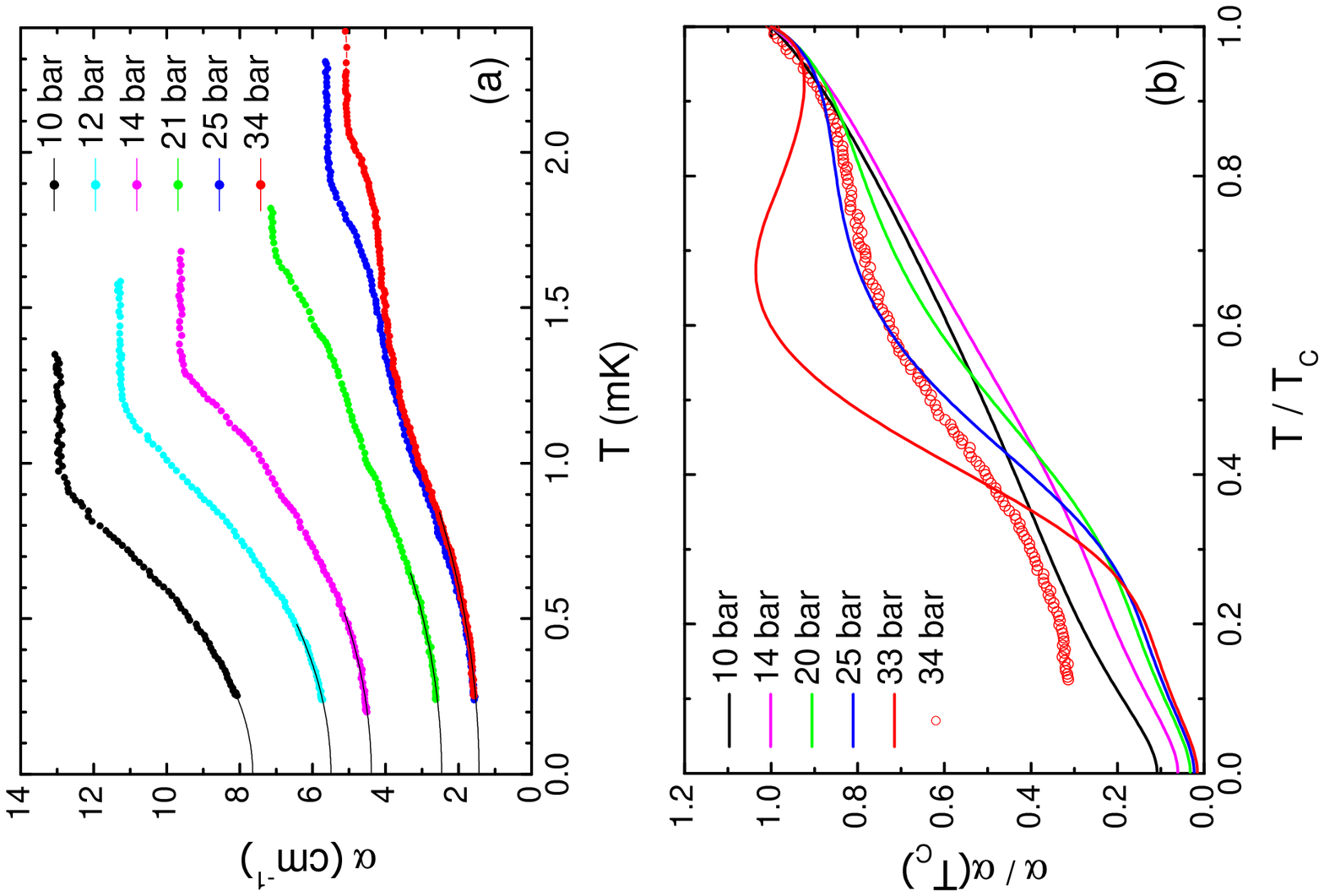}
\caption{\label{fig:epsart}(a) Absolute attenuation for various
pressures vs.\, temperature (color in on-line version). Thin solid
lines are the results of a quadratic fit to the low temperature
part ($T/T_{c}\lesssim$~0.4) of the data at each pressure. (b)
Normalized sound attenuation vs. normalized temperature. The
results of theoretical calculation (solid lines, color in on-line
version) are plotted along with the experimental results at 34
bars for comparison.}
\end{figure}

Low mass density and the compliant nature of aerogel necessitate
the consideration of effective momentum transfer upon
quasi-particle scattering off the aerogel, which generates dragged
motion of aerogel. Ichikawa {\it et al.}\,\cite{ich} incorporated
the collision drag effect in calculating the dispersion relation
in the normal fluid. Their model offered a successful explanation
for the experimental results of the Northwestern group \cite{nom}.
Recently, Higashitani {\it et al.}\,\cite{miu,hig} extended this
model to study the longitudinal sound (fast mode) propagation in
superfluid $^3$He/aerogel within the framework of the two-fluid
model. The drag effect can be described phenomenologically by a
frictional force,
$\vec{F}_{d}={\rho_{n}\over{\tau_{f}}}(\vec{v}_{n}-\vec{v}_{a})$,
introducing an additional relaxation time $\tau_{f}$, where
$\vec{v}_{n(a)}$ is the normal fluid component (aerogel) velocity.
This effect is of particular importance when $\omega\tau_{i}<$~1,
and the total attenuation (Eq.~(130) of ref.~\cite{hig}) is
\begin{equation}
{\alpha} =
{{\omega^{2}/2c_{f}\over{1+\rho_{a}\rho_{s}/\rho_{n}\rho}}
({{\rho_{a}^{2}\tau_{f}/\rho\rho_{n}}\over{1+\rho_{a}/\rho_{n}}}+
{{4\eta/3\rho c_{1}^{2}}\over{1+\rho_{a}\rho_{s}/\rho_{n}\rho}})},
\end{equation}
where $\eta$ is the shear viscosity of liquid $^3$He.  The first
term ($\alpha_{f}$) arises from the frictional damping caused by
the aerogel motion relative to the normal fluid component, and the
second term ($\alpha_{v}$) from the conventional hydrodynamic
sound damping associated with the viscosity. This expression
allows us to extract $\ell$ in this system from our absolute
attenuation at the transition temperature, $\alpha_{c}$. The inset
of Fig.~3 shows our results of $\alpha_{c}$ for various pressures.
The solid lines are the result of calculation using Eq.~(1) for
three different mean free paths, $\ell=$~100, 120, and 140~nm.  As
can be seen, $\ell=$~120~nm produces an excellent fit to our data
for the whole pressure range, which is in good agreement with the
values obtained from the thermal conductivity (90 nm) \cite{fis}
and spin diffusion (130~nm) \cite{sau} measurements. With the
knowledge of the mean free path, one can calculate the full
temperature dependence of sound attenuation in the superfluid
phase. The results of the calculation (in the unitary limit)
following the prescription described in ref.\,\cite{hig} are
displayed in Fig.~2(b) along with the experimental results at 34
bars. The calculation reproduces all the important features
observed in our measurements.  In particular, the conspicuous
shoulder structure appearing near $T/T_{c}\approx$~0.6 at 33 bars
softens at lower pressures and is completely absorbed in an almost
linear temperature dependence below 20 bars.  This behavior is the
characteristic of $\alpha_{f}$ \cite{hig}. A fast decrease in
$\rho_{n}$ right below $T_{c}$ produces the bump in $\alpha_{f}$,
and $\alpha_{f}\rightarrow$~0 as $T\rightarrow$~0. On the other
hand, $\alpha_{v}$ decreases monotonically and reaches a finite
value due to non-zero $\rho_{n}$ and the impurity states induced
inside the gap as T $\rightarrow$ 0.  The quantitative agreement
between the theory and experiment, however, is not yet
satisfactory.  The calculation utilizes the isotropic homogeneous
scattering model (IHSM) \cite{thu}, which tends to overestimate
$\Delta(T)$ and $\rho_{s}$ compared to the experimentally
determined values \cite{por,gol}. As shown in ref.~\cite{han}, the
inhomogeneity gives rise to the reduction of the average value of
the order parameter and consequently yields larger $\eta$ and
$\rho_{n}$, which in turn increases $\alpha_{0}$ but decreases the
frictional contribution. It is also expected that the non {\it
s-wave} scattering components  make non-trivial contributions to
the viscous and frictional relaxation times in a direction that
improves the quantitative agreement.

\begin{figure}
\includegraphics[height=2.75in, angle=-90]{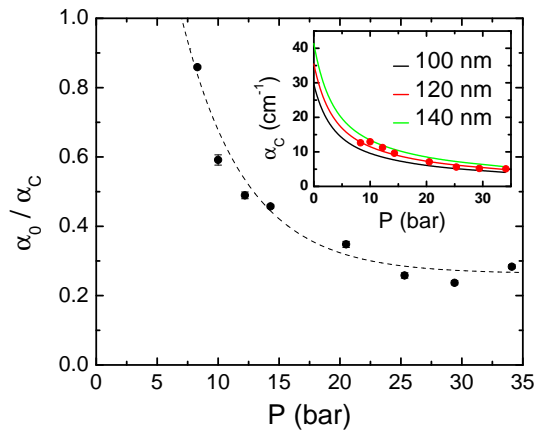}
\caption{\label{fig:epsart} Normalized zero temperature
attenuation vs.\, pressure. The dashed line is a guide for eye.
Inset: Pressure dependence of sound attenuation at $T_{c}$. The
solid lines (color on-line) are the results of theoretical fit for
$\ell =$~100, 120, and 140~nm (see text).}
\end{figure}

Theoretical calculations based on the IHSM \cite{sha,hig} predict
that the impurity states would completely fill the gap, leading to
a gapless superfluid when $\tau_{i}T_{c}<$~1 for the B-phase in
the unitary limit.  We estimate 0.3~$<\tau_{i}T_{c}<$~1 for 10~$<P
<$~34 bars with $\ell=$~120~nm.  The normalized zero temperature
attenuation ($\alpha_{0}/\alpha_{c}$) obtained by extrapolating
the low temperature part of the attenuation (solid lines in
Fig.~2(a)) is plotted in Fig.~3, where $\alpha_{0}/\alpha_{c}$
increases as the sample pressure is reduced and seems to approach
unity near $P_{c}\approx$~6 bars. Since the viscosity ratio is
directly related to the density of states at zero energy through
$\eta(0)/\eta(T_{c}) = n(0)^{z}$, $z =$~\{2,4\} for the \{Born,
unitary\} limit where $n(0)$ is the normalized density of states
at zero energy \cite{hig}, the finite $\alpha_{0}/\alpha_{c}$ is
strong evidence of a finite $n(0)$. The gapless behavior has been
experimentally suggested by recent thermal conductivity (for
$P\leq$~10 bars) \cite{fis} and heat capacity (for 11~$\leq
P\leq$~29 bars) \cite{choNW} measurements. The pressure dependence
of $\alpha_{0}/\alpha_{c}$ is in qualitative agreement with the
combined results of Fisher {\it et al.}\, and Choi {\it et al.}
Although all of these experimental techniques (including ours) are
limited to probe the impurity states near the Fermi level, the
behavior is consistent with the theoretical predictions with
gapless excitations. Unlike the thermodynamic and transport
measurements, the high frequency ultrasound measurement has a
potential to unveil a larger portion of the impurity states
profile from the frequency dependence.


We acknowledge support from an Alfred P. Sloan Research Fellowship
(YL), NSF grants DMR-0239483 (YL), DMR-0305371 (MWM), and a
Grant-in-Aid for Scientific Research on Priority Areas (No.
17071009) from MEXT of Japan (SH and KN). We would like to thank
J.-H. Park for his technical assistance, and Jim Sauls, Peter
W\"olfle, and Bill Halperin for useful discussions.


\begin{thebibliography}{30}
\bibitem{com1} For example, $c=$~350~($\pm$ 10)~m/s at 34 bars from our measurement,
and $c_{f}=$~370~m/s.
\end{thebibliography}
\end{document}